
\input phyzzx
\font\elevenmib=cmmib10 scaled\magstephalf   \skewchar\elevenmib='177
\catcode`\@=11 
\def\KUNSmark{\vtop{\hbox{\elevenmib Department\hskip1mm of\hskip1mm
             Physics}\hbox{\elevenmib Kyoto\hskip1mm University}}}
\newif\ifKUNS \KUNStrue
\def\titlepage{\FRONTPAGE\papers\ifPhysRev\PH@SR@V\fi
    \ifKUNS\null\vskip-17mm\KUNSmark\vskip0mm\fi
    \ifp@bblock\p@bblock \else\hrule height\z@ \rel@x \fi }
\catcode`\@=12 
\Pubnum={KUNS 1246}
\date={February 1994}
\titlepage
\title{Post-Newtonian Expansion of the
       Ingoing-Wave Regge-Wheeler Function}
\author{Misao SASAKI}
\address{Department of Physics, Faculty of Science,\break
Kyoto University, Kyoto 606}
\bigskip\medskip
\centerline{Submitted to \sl Prog. Theor. Phys.}
\abstract{
We present a method of post-Newtonian expansion to
solve the homogeneous Regge-Wheeler equation which describes
gravitational waves on the Schwarzschild spacetime.
The advantage of our method is that it allows a systematic
iterative analysis of the solution.
Then we obtain the Regge-Wheeler function
which is purely ingoing at the horizon in closed analytic form,
with accuracy required to determine the gravitational wave luminosity
to (post)$^{4}$-Newtonian order (i.e., order $v^8$ beyond Newtonian)
from a particle orbiting around a Schwarzschild black hole.
Our result, valid in the small-mass limit of one body,
gives an important guideline for the study of coalescing
compact binaries.
In particular, it provides basic formulas to analytically calculate
detailed waveforms and luminosity,
including the tail terms to (post)$^3$-Newtonian order,
which should be reproduced in any other post-Newtonian calculations.
}
\def\Buildrel#1\under#2{\mathrel{\mathop{#2}\limits_{#1}}}
\def\Ci{\mathop{\rm Ci}\nolimits}
\def\Si{\mathop{\rm Si}\nolimits}
\def\ci{\mathop{\rm ci}\nolimits}
\def\si{\mathop{\rm si}\nolimits}
\def\CC{{C}}
\def\SS{{S}}
\def\Im{\mathop{\rm Im}}
\def\Re{\mathop{\rm Re}}
\REF\LIGO{A. Abramovici et al., Science {\bf 256} (1992) 325.}
\REF\VIRGO{C. Bradaschia et al., Nucl. Instrum. \& Methods
 {\bf A289} (1990) 518.}
\REF\Thorne{see e.g., K.S. Thorne, in {\it Proceedings of the 8th
 Nishinomiya-Yukawa Symposium on Relativistic Cosmology}, ed.
 M. Sasaki (Universal Academy Press, Tokyo), to appear.}
\REF\Will{see e.g., C.M. Will, in {\it Proceedings of the 8th
 Nishinomiya-Yukawa Symposium on Relativistic Cosmology}, ed.
 M. Sasaki (Universal Academy Press, Tokyo), to appear.}
\REF\Cutler{C. Cutler et al., Phys. Rev. Lett. {\bf 70} (1993) 2984.}
\REF\TagNak{H. Tagoshi and T. Nakamura,
  Preprint KUNS 1223, to be published in Phys. Rev. D.}
\REF\RegWhe{T. Regge and J.A. Wheeler, Phys. Rev.
 {\bf 108} (1957) 1063.}
\REF\Poisson{E. Poisson, Phys. Rev. D {\bf 47} (1993) 1497.}
\REF\Teukol{S.A. Teukolsky, Astrophys. J. {\bf 185} (1973) 635.}
\REF\Chandra{S. Chandrasekhar,
   Proc. R. Soc. London {\bf A343} (1975) 289.}
\REF\SasNak{M. Sasaki and T. Nakamura,
   Phys. Lett. {\bf 87A} (1981) 85.}
\REF\Tanaka{T. Tanaka, M. Shibata, M. Sasaki, H. Tagoshi and
   T. Nakamura, Prog. Theor. Phys. {\bf 90} (1993) 65.}
\REF\Apoetal{T. Apostolatos, D. Kennefick, A. Ori and E. Poisson,
  Phys. Rev. D {\bf 47} (1993) 5376.}
\REF\Wiseman{A.G. Wiseman, Phys. Rev. D {\bf 48} (1993) 4757.}
\REF\TagSas{H. Tagoshi and M. Sasaki, in preparation.}
\chapter{Introduction}

Among the possible sources of gravitational radiation,
coalescing compact binaries have become one of the most
promising candidates which may
be detected by the near-future laser interferometric gravitational
wave detectors such as LIGO\refmark{\LIGO} or
VIRGO\rlap.\refmark\VIRGO\
Among various others, one very crucial reason for this is
because of the simplicity of physics involved during the orbital
evolution of such a system.
Except for the very last moment of the coalescence,
the dynamics is dominated by gravity alone and is almost
non-relativistic. The energy carried away by gravitational waves
per orbital period is small enough and the orbit spirals in very
slowly. Thus the gravitational radiation from the inspiraling binary
has a distinguishable waveform which may be easily detected.
Then by matched filtering\rlap,\refmark\Thorne we may be able to
determine the orbital parameters, spins of the constituent stars,
the distance to the source, etc..
Astrophysical importance of such measurements is needless to say.
Consequently much effort has been recently made to the study of the
inspiral stage of coalescing compact binaries\rlap.\refmark{\Will}\

However, based on calculations of the gravitational radiation from
a particle in circular orbit around a non-rotating black hole,
which is valid when one body of the binary
has mass much smaller than the other and each has no spin,
Cutler et al\rlap.\refmark{\Cutler} showed that
evaluation of the gravitational wave luminosity to a post-Newtonian
order much higher than what can be presently achieved will be
necessary to construct templates effective enough to determine
the source parameters.
Then the same problem was studied with much greater accuracy
by Tagoshi and Nakamura\refmark{\TagNak} in a semi-analytic way.
In this very interesting work they calculated the coefficients of
post-Newtonian expansion of the gravitational wave luminosity to
(post)$^4$-Newtonian order (i.e., $O(v^{8})$ beyond Newtonian)
and claimed that the accuracy to at least (post)$^3$-Newtonian order
is required for the construction
of effective theoretical templates. In addition, they found
logarithmic terms in the luminosity formula at and beyond
(post)$^3$-Newtonian order.

It is then a matter of great importance to confirm those coefficients
they calculated and to understand the origin of logarithmic terms.
In this paper, we present a method to systematically calculate
the gravitational radiation in the test particle limit and
show that an analytic derivation of those coefficients
is indeed possible. Our method is an extension of the one originally
developed by Poisson\rlap,\refmark{\Poisson} in which the expansion
parameter is $M\omega$ where $M$ is the mass of a Schwarzschild
 black hole (or a non-rotating compact star) and $\omega$ is the
frequency of gravitational waves, but in a way which is
more suited for a systematic post-Newtonian expansion.
In particular, we find it can more adequately deal with the
logarithmic ambiguity one encounters when one matches light cones
in the near zone and far zone in the post-Newtonian expansion.

The paper is organized as follows. In \S2, we formulate
a new post-Newtonian approach to the
gravitational radiation emitted by a particle orbiting
around a non-rotating black hole, which is governed by the
Regge-Wheeler equation\rlap.\refmark\RegWhe\
We then give an iterative method to
solve for the Regge-Wheeler function, $X^{in}_{\omega\ell}(r)$,
describing a purely ingoing wave at the black-hole horizon,
which plays an essential role in evaluating the gravitational
radiation from a particle.
In \S3, as a preliminary, we analyze general features of the
iterative solution of $X^{in}_{\omega\ell}$. We clarify the required
accuracy of $X^{in}_{\omega\ell}$ in order to evaluate
the gravitational wave luminosity and waveforms to $O(v^8)$
beyond Newtonian. Namely, we find the calculation to
$O\left((M\omega)^2\right)$ is necessary only for $\ell\leq3$
to derive the luminosity formula to $O(v^8)$,
 but additionally for $\ell=4$ to derive the waveform.
In \S4, we give the solution to $O(M\omega)$ for arbitrary $\ell$
and to $O\left((M\omega)^2\right)$ for $\ell\leq4$
in closed analytic form.
Finally \S5 is devoted to conclusion. Some useful mathematical
formulas used in the text are given in Appendix.
Throughout the paper we use geometrized units, $c=G=1$.

\chapter{Formulation}

The situation we consider is the limit in which one of compact binary
objects has a small mass relative to the other and each
has no spin. In this case, the larger can be approximated
by a non-rotating black hole (with mass $M$) and the smaller can be
treated as a test particle (with mass $\mu\ll M$) oribiting around it.
Of course, except for the case the larger body is actually a
non-rotating black hole, approximating it by a black hole fails when
the orbit approaches the event horizon. However, as long as one
is interested in the regime in which one can neglect the radiation
going into the black hole, this should be a valid approximation.
In fact, we will see below that as long as we focus on the
radiation outgoing to infinity, it is so until we reach
$O(v^{18})$ beyond Newtonian.

It is known that the gravitational radiation from such a system is
described by the Teukolsky equation\rlap,\refmark{\Teukol}
 but it is also known that it can be transformed to
the Regge-Wheeler equation\rlap,\refmark{\Chandra,\SasNak} which has
a much simpler form.
Here we consider the Regge-Wheeler equation.
The inhomogeneous Regge-Wheeler equation takes the form,
$$
\left[{d^2\over dr^{*2}}+\omega^2
      -\left(1-{2M\over r}\right)
        \left({\ell(\ell+1)\over r^2}-{6M\over r^3}\right)
     \right]X_{\omega\ell m}=\mu S_{\omega\ell m}\,,
\eqn\RWeq
$$
where $S_{\omega\ell m}$ is the source term determined from the
orbit of the particle and
$r^*=r+2M\ln(r/2M-1)$ is the tortoise coordinate.
The explicit form of $S_{\omega\ell m}$ for bound orbits
can be found in Ref.\Tanaka).

The solution $X_{\omega\ell m}$ which satisfies the outgoing-wave
condition at infinity and ingoing-wave condition at horizon is
then given by
$$
\eqalign{
 X_{\omega\ell m}(r)&={\mu\over W}
      \int_{-\infty}^\infty dr^{*'}G(r,r')S_{\omega\ell m}(r')\,;
\crr
 &G(r,r')=\theta(r-r')X^{out}(r)X^{in}(r')+
         \theta(r'-r)X^{in}(r)X^{out}(r')\,,
\cr
 &W=X^{in}{d\over dr^*}X^{out}-X^{out}{d\over dr^*}X^{in}\,,
\cr}
\eqn\Xradsol
$$
where $X^{in}$ and $X^{out}$ are the solutions of the homogeneous
Regge-Wheeler equation with the boundary conditions,
$$
\eqalign{
 &X^{in}\rightarrow
  \cases{Ce^{-i\omega r^*}~&
                      for $r^*\rightarrow-\infty,$\cr
         A^{in}e^{-i\omega r^*}+A^{out}e^{i\omega r^*}~&
                      for $r^*\rightarrow\infty,$\cr}\,
\cr
 &X^{out}\rightarrow
  \cases{B^{in}e^{-i\omega r^*}+B^{out}e^{i\omega r^*}~&
                      for $r^*\rightarrow-\infty,$\cr
         e^{i\omega r^*}~&
                      for $r^*\rightarrow\infty,$\cr}\,
\cr}
\eqn\Xinout
$$
respectively.
Note that from the constancy of the Wronskian for any two solutions,
we have
$$
\eqalign{
 &|A^{in}|^2-|A^{out}|^2=|C|^2,\qquad |B^{out}|^2-|B^{in}|^2=1,
\cr
 & W=2i\omega A^{in}=2i\omega CB^{out}.
\cr}
\eqn\Wronskian
$$
Usually one normalizes $X^{in}$ so that $C=1$,
but we leave it arbitrary here for later convenience.

Now as long as we are interested only in the radiation emitted
to infinity, it is clear from Eq.\Xradsol\ that we only
need to know the form of $X^{in}$. That is,
we have
$$
 X_{\omega\ell m}(r)\rightarrow \mu A_{\omega\ell m}
                   e^{i\omega r^*};
\quad
 A_{\omega\ell m}
    ={\int_{-\infty}^\infty dr^*X^{in}(r)S_{\omega\ell m}(r)
            \over 2i\omega A^{in}}\,,
\eqn\Xinfty
$$
as $r\rightarrow\infty$.
Then the wave form at infinity is expressed as
$$
 h_{+}-ih_{\times}\sim {8\mu\over r}
   \sum_{\ell m}\int d\omega A_{\omega\ell m}
         e^{-i\omega (t-r^*)}\,{}_{-2}Y_{\ell m}(\theta,\varphi)\,,
\eqn\waveform
$$
where ${}_{-2}Y_{\ell m}$ is the spherical harmonic
function of spin weight $s=-2$. From the above waveform,
it is straightforward to calculate
the gravitational wave luminosity.
For a circular orbit, the gravitational
radiation spectrum is discrete:
$$
 A_{\omega\ell m}=\delta(\omega-m\Omega)\tilde A_{\ell m}\,,
\eqn\circular
$$
where $\Omega$ is the angular velocity.
In this case, the gravitational wave luminosity
is given by
$$
 {dE\over dt}={1\over16\pi}\int
    \left(|\dot h_{+}|^2+|\dot h_{\times}|^2\right)r^2d\Omega
    =\sum_{\ell m}
        {16\mu^2\omega^2\left|\tilde A_{\ell m}\right|^2\over4\pi}\,.
\eqn\dEdt
$$

In order to formulate a post-Newtonian expansion, we rewrite
the homogeneous Regge-Wheeler equation as
$$
  \left[{d^2\over dz^{*2}}+1-
         \left(1-{\epsilon\over z}\right)
      \left({\ell(\ell+1)\over z^2}-{3\epsilon\over z^3}\right)
     \right]X^{in}_\ell=0,
\eqn\HRWeq
$$
where $z=\omega r$, $z^*=z+\epsilon\ln(z-\epsilon)$
and $\epsilon=2M\omega$, and
hereafter we attach the eigenvalue index $\ell$ to $X^{in}$
for clarity.
In the case of a circular orbit at radius $r=r_0$,
we have $\Omega r_0=(M/r_0)^{1/2}=v$ and $\omega =m\Omega$,
where $v$ is the orbital velocity.
Hence $z_0=\omega r_0\sim v$ and $M\omega=(M/r_0)r_0\omega\sim v^3$.
Further the source term $S_{\omega\ell m}$ has support only for
$z\leq z_0$\rlap.\refmark{\Tanaka}%
\footnote{\dag}
{One may also transform $X^{in}_\ell$ to the corresponding
ingoing-wave Teukolsky function first and then integrate the
inhomogeneous Teukolsky equation.
In this case, the source term has support only at
$z=z_0$\rlap.\refmark\Poisson }
Hence the post-Newtonian expansion corresponds to solving \HRWeq\
recursively with respect to $\epsilon$ and then expanding the
resulting $X^{in}_\ell$ in powers of $z$ as well as to extract out
its incident amplitude $A^{in}_\ell$ at infinity to required
accuracy.

Now, since $z^*$ contains $\epsilon$ in itself, a naive expansion
of \HRWeq\ would result in a complicated equation which involves
many $\epsilon$-dependent terms. Furthermore, such an expansion is not
quite desirable at both boundaries since $X^{in}_\ell\sim e^{-iz^*}$
at horizon ($z^*\rightarrow-\infty$) and
$X^{in}_\ell\sim e^{\pm iz^*}$ at infinity ($z^*\rightarrow\infty$).
In order to take into account these asymptotic behaviors at both
boundaries, one would have to take $z^*$ as the independent variable.
However, this is not desirable either since the post-Newtonian
expansion assumes an infinitesimally small $\epsilon$, but $z^*$ is not
analytic at $\epsilon\rightarrow0$ ($z^*\rightarrow-\infty$
as $z\rightarrow\epsilon$, while taking the limit
$\epsilon\rightarrow0$ first yields $z^*\rightarrow z$ which never
approaches $-\infty$). One possibility to circumvent this non-analytic
behavior is to separate out $e^{-iz^*}$ dependence from the
dependent variable $X^{in}_\ell$ in the beginning and to adopt $z$
as the independent variable. Incidentally, this also takes
care of the incoming part of the asymptotic behavior at infinity.
In a sense, the incoming light-cone is adequately respected
in this way.
It turns out this idea works pretty well. Setting
$$
 X^{in}_\ell=e^{-i\epsilon\ln(z-\epsilon)}z\xi_\ell(z),
\eqn\Xinform
$$
we find \HRWeq\ reduces to the form,
$$
     \left[{d^2\over dz^2}+{2\over z}{d\over dz}
       +\left(1-{\ell(\ell+1)\over z^2}\right)\right]\xi_\ell(z)
      =\epsilon e^{-iz}{d\over dz}\left[{1\over z^3}
          {d\over dz}\left(e^{iz}z^2\xi_\ell(z)\right)\right].
\eqno\eq
$$
Surprisingly, $\epsilon$ appears only as an overall factor on
the right-hand side but nowhere else. Hence it readily
allows us to apply the expansion in terms of $\epsilon$:
$$
  \xi_\ell(z)=\sum_{n=0}^{\infty}\epsilon^n \xi_\ell^{(n)}(z)\,;
\eqn\xiexpand
$$
for which we have
$$
\eqalign{
  &\left[{d^2\over dz^2}+{2\over z}{d\over dz}
  +\left(1-{\ell(\ell+1)\over z^2}\right)\right]\xi_\ell^{(n)}(z)
\crr
   &\hskip20mm   =e^{-iz}{d\over dz}\left[{1\over z^3}
         {d\over dz}\left(e^{iz}z^2\xi_\ell^{(n-1)}(z)\right)\right].
\cr}
\eqn\PNexpand
$$

Once $\xi_\ell$ is obtained to desired accuracy,
$X^{in}_\ell$ is readily obtained by inserting the resulting
$\xi_\ell$ into Eq.\Xinform. The calculation of $A^{in}_\ell$ is
then straightforward by gathering together all the coefficients
of leading terms proportional to $e^{-iz}/z$ from
the asymptotic form of $\xi_\ell$ at $z=\infty$.
In the rest of this section, we describe
a method to carry out this procedure.

Solving Eq.\PNexpand\ recursively is in principle straightforward.
Since the general homogeneous solution to the left-hand side of it
is given by a linear combination of the spherical Bessel functions
$j_\ell$ and $n_\ell$, one can immediately write the integral
expression for $\xi^{(n)}_\ell$. Noting that
 $j_\ell n_\ell'-n_\ell j_\ell'=1/z^2$, we have
$$
\eqalign{
 \xi_\ell^{(n)}=&n_\ell\int dz\,z^2e^{-iz}j_\ell
      \left[{1\over z^3}
         \left(e^{iz}z^2\xi^{(n-1)}(z)\right)'\right]'
\crr
 &-j_\ell\int dz\,z^2e^{-iz}n_\ell
      \left[{1\over z^3}
         \left(e^{iz}z^2\xi^{(n-1)}(z)\right)'\right]',
\cr}
\eqn\Indefform
$$
where the prime denotes $d/dz$ and we have tentatively expressed
the integrals in the indefinite form for convenience.
If these indefinite integrals can be explicitly performed,
the boundary condition for $X^{in}_\ell$ as given in Eq.\Xinout\
is easily implemented.
For this purpose, we first rewrite Eq.\Indefform\
by performing integration by parts as
$$
\eqalign{
 \xi_\ell^{(n)}
  &=-n_\ell\int dz\,\left(z^2e^{-iz}j_\ell\right)'{1\over z^3}
             \left(z^2e^{iz}\xi_\ell^{(n-1)}\right)'
      +j_\ell\int zdz\,\left[j_\ell\rightarrow n_\ell\right]
\crr
  &=-n_\ell\int dz\,z\Biggl\{j_\ell'\xi_\ell^{(n-1)}{}'
           +{2\over z}(j_\ell\xi_\ell^{(n-1)})'
           +\left({4\over z^2}+1\right)j_\ell\xi_\ell^{(n-1)}
\crr
  &\quad\qquad
   -i\left(j_\ell\xi_\ell^{(n-1)}{}'-j_\ell'\xi_\ell^{(n-1)}\right)
           \Biggr\}
          +j_\ell\int dz\,\left[j_\ell\rightarrow n_\ell\right]
\crr
  &=n_\ell\int dz\,\left\{\left((zj_\ell')'
                         -\left(z+{4\over z}\right)j_\ell
                         \right)\xi_\ell^{(n-1)}
   +iz\left(j_\ell\xi_\ell^{(n-1)}{}'-j_\ell'\xi_\ell^{(n-1)}\right)
                \right\}
\crr
   &\quad\qquad-j_\ell\int dz\,\left[j_\ell\rightarrow n_\ell\right]
      +z(j_\ell n_\ell'-n_\ell j_\ell')\xi_\ell^{(n-1)}.
\cr
}\eqno\eq
$$
Then using recurrence formulas
for the spherical Bessel functions, the above is further
rewritten as
$$
\eqalign{
 &\xi_\ell^{(n)}
    ={1\over z}\xi_\ell^{(n-1)}
          +j_\ell I_{\ell,n}^{(n-1)}-n_\ell I_{\ell,j}^{(n-1)}
\crr
 &~+in_\ell\int dz\,z
     \left(j_\ell\xi_\ell^{(n-1)}{}'-j_\ell'\xi_\ell^{(n-1)}\right)
   -ij_\ell\int dz\,z
     \left(n_\ell\xi_\ell^{(n-1)}{}'-n_\ell'\xi_\ell^{(n-1)}\right),
\cr}
\eqn\Genform
$$
where $I_{\ell,\zeta}^{(n)}$ is an indefinite integral defined as
$$
 I_{\ell,\zeta}^{(n)}:=
 \int dz\left\{
     2z{\zeta}_\ell-{(\ell-1)(\ell+3)\over2\ell+1}{\zeta}_{\ell+1}
     -{\ell^2-4\over2\ell+1}{\zeta}_{\ell-1}\right\}\xi^{(n)}_\ell\,.
\eqn\Ielldef
$$
Here ${\zeta}_\ell$ stands for either $j_\ell$ or $n_\ell$.

Now let us consider the boundary conditions for $\xi_\ell^{(n)}$.
The boundary condition for $X^{in}_\ell$ as
$z^*\rightarrow-\infty$ $(z\rightarrow\epsilon)$
is that $X^{in}_\ell=e^{-\epsilon\ln(z-\epsilon)}z\xi_\ell
\propto e^{-iz^*}$. Hence $z\xi_\ell$ is regular at
$z=\epsilon$. Since $\epsilon$ can be taken arbitrarily small,
this implies that $z\xi_\ell^{(n)}$ must be no more
singular than $O(z^{-n})$ at $z=0$.
In particular, this fixes the lowest order solution to be
$\xi^{(0)}_\ell=N_\ell j_\ell$ where $N_\ell$ is an
arbitrary normalization constant. For convenience, we set
$N_\ell=1$. Taking account of the behavior of the lowest order
 solution and the structure of the basic equation \PNexpand,
 we then infer that $z\xi^{(n)}_\ell$ must be no more singular
than $z^{\ell+1-n}$ at $z=0$. Thus for $n\leq3$, which is sufficient
for the purpose of this paper, the boundary
condition is that all the $z\xi^{(n)}_\ell$ are regular at $z=0$.
In passing, we note a symmetry of $\xi_\ell$. From the form of
Eq.\PNexpand\ and the fact that $\xi^{(0)}_\ell=j_\ell$ and
$j_\ell(-z)=(-1)^\ell j_\ell(z)$, we deduce that
$$
 \xi^{(n)}_\ell(-z)=(-1)^{\ell+n}\overline{\xi^{(n)}_\ell(z)}\,,
\eqn\xisym
$$
where the bar denotes complex conjugation.
It is worthwhile to keep this symmetry in mind.

 Since $j_\ell=O(z^\ell)$ as $z\rightarrow0$, we have
$X^{in}_\ell\rightarrow O(\epsilon^{\ell+1})e^{-iz^*}$,
or $C_\ell=O(\epsilon^{\ell+1})$ in the notation of Eq.\Xinout.
On the other hand, from the asymptotic behavior of $j_\ell$ at
$z=\infty$, we find $A^{in}_\ell$ and $A^{out}_\ell$ are
of order unity. Then from the first of Eq.\Wronskian,
we obtain
$$
 |A^{in}_\ell|-|A^{out}_\ell|
   ={|C_\ell|^2\over|A^{in}_\ell|+|A^{out}_\ell|}
    =O(\epsilon^{2\ell+2}).
\eqno\eq
$$
Thus $|A^{in}_\ell|=|A^{out}_\ell|$ until we go to
$O(\epsilon^{2\ell+2})$ or more.
Since $\ell\geq2$, this implies that even though we are
dealing with the Regge-Wheeler equation which describes gravitational
radiation from a non-rotating black hole, it also correctly describes
a situation with a general non-rotating compact star,
provided that the inner
structure of the star may be ignored and we focus on the outgoing
radiation up to the amplitude of less than $O(\epsilon^{6})$ or
$O(v^{18})$ beyond Newtonian.
We note that
if one considers the effect of radiation reaction to the
orbital evolution, difference appears between the cases of
a black hole and a star much earlier, at $O(v^8)$
beyond Newtonian, due to the radiation absorbed into the
black-hole horizon\rlap.\refmark{\Apoetal}\
What we claim here is that the effect
of the presence of the horizon on the {\it outgoing\/} radiation
appears at $O(v^{18})$.

The above fact also implies that we may choose the phase
of $X^{in}_\ell$ so that $A^{in}_\ell$ and $A^{out}_\ell$ are
complex conjugate to each other. Then since the Regge-Wheeler equation
is real, it follows that $X^{in}_\ell$ can be made real up to
$O(\epsilon^5)$.
Provided we choose the phase of $X^{in}_\ell$ in this way,
$\Im\left(\xi^{(n)}_\ell\right)$ for a given $n$
is completely determined in terms of $\Re\left(\xi^{(r)}_\ell\right)$
for $r\leq n-1$.
To see this, let us decompose
the real and imaginary parts of $\xi^{(n)}_\ell$:
$$
 \xi^{(n)}_\ell=f^{(n)}_\ell+ig^{(n)}_\ell\,.
\eqn\fgdef
$$
Inserting this expression into Eq.\Xinform\ and expanding the result
with respect to $\epsilon$ by assuming $z\gg\epsilon$, we find
$$
\eqalign{
 X^{in}_\ell
     =&e^{-i\epsilon\ln(z-\epsilon)}z
         \left(j_\ell+\epsilon(f^{(1)}_\ell+ig^{(1)}_\ell)
             +\epsilon^2(f^{(2)}_\ell+ig^{(2)}_\ell)+\cdots\right)
\crr
    =&z\left(j_\ell+\epsilon f^{(1)}_\ell
        +\epsilon^2\Bigl(f^{(2)}_\ell+g^{(1)}_\ell\ln z
        -{1\over2}j_\ell(\ln z)^2\Bigr)+\cdots\right)
\cr
     &+iz\left(\epsilon(g^{(1)}_\ell-j_\ell\ln z)
           +\epsilon^2\Bigl(g^{(2)}_\ell+{1\over z}j_\ell
                         -f^{(1)}_\ell\ln z\Bigr)+\cdots\right).
\cr}
\eqn\Xinexpand
$$
Hence we must have
$$
g^{(1)}_\ell=j_\ell\ln z\,,\quad
g^{(2)}_\ell=-{1\over z}j_\ell+f^{(1)}_\ell\ln z\,,\quad\cdots.
\eqn\Imxi
$$
Since the reality of $X^{in}_\ell$ is not manifest in our
expansion scheme,
this fact can be used as a useful method of checking calculations.
To conclude this section, we note the relation between the
functions $f^{(n)}_\ell$ and
the conventional post-Newtonian expansion of $X^{in}_\ell$:
$$
\eqalign{
   &X^{in}_\ell=\sum_{n=0}^{\infty}\epsilon^n X^{(n)}_\ell\,;
\crr
   &X^{(0)}_\ell=zf^{(0)}_\ell=zj_\ell\,,\quad
    X^{(1)}_\ell=zf^{(1)}_\ell\,,\quad
    X^{(2)}_\ell=z\left(f^{(2)}_\ell+{1\over2}j_\ell(\ln z)^2\right),
   \quad\cdots.
\cr}
\eqn\Xxirel
$$
\chapter{Preliminary analysis}
Prior to actually carrying out the integration of Eq.\Genform,
in this section we investigate general features of
the asymptotic behavior of $\xi^{(n)}_\ell$ at $z=0$ and $z=\infty$
for $n=1$ and $2$. In addition, we comment on the behavior of
$\xi^{(3)}_\ell$ at $z=0$ for $\ell\leq4$.
Consequently, we clarify the condition to achieve $O(v^8)$
beyond Newtonian.

First we consider the asymptotic behavior at $z=0$.
We return to Eq.\PNexpand\ and represent the right-hand side of it
by $S^{(n)}_\ell$. As we mentioned at the end of the previous section,
the imaginary parts of $\xi^{(n)}_\ell$ are subject to their real parts
of lower $n$. Hence we may focus on the real parts.

For $n=1$, we have
$$
\eqalign{
 \Re\left(S^{(1)}_\ell\right)
    &={1\over z}\left(j_\ell''+{1\over z}j_\ell'
                       -{4+z^2\over z^2}j_\ell\right)
\cr
    &=\cases{O(z)~&for $\ell=2$,\cr
             O(z^{\ell-3})~&for $\ell\geq3$.\cr}
\cr}
\eqn\Szero
$$
Hence from Eq.\PNexpand\ or \Indefform\ and noting the regularity
at $z=0$, we obtain
$$
 \Re\left(\xi^{(1)}_\ell\right)=f^{(1)}_\ell
   =\cases{O(z^3)+\alpha^{(1)}_2j_2(z)~&for $\ell=2$,\cr
           O(z^{\ell-1})
              +\alpha^{(1)}_\ell j_\ell(z)~&for $\ell\geq3$,\cr}
\eqn\Rexione
$$
where $\alpha^{(1)}_\ell$ is an arbitrary constant. Since its presence
simply corresponds to a different normalization of $X^{in}_\ell$,
we set $\alpha^{(1)}_\ell=0$ for simplicity.

For $n=2$, we then have
$$
\eqalign{
 \Re\left(S^{(2)}_\ell\right)
    &={1\over z}\left(f^{(1)}_\ell{}''+{1\over z}f^{(1)}_\ell{}'
                       -{4+z^2\over z^2}f^{(1)}_\ell\right)
      -{1\over z}\left(2g^{(1)}_\ell{}'+{1\over z}g^{(1)}_\ell\right)
\crr
    &=-{1\over z}\left(j_\ell\ln z\right)'-{1\over z^2}j_\ell\ln z
       +\cases{O(z^{\ell-2})~&for $\ell=2,3$,\cr
               O(z^{\ell-4})~&for $\ell\geq4$,\cr}
\cr}
\eqn\Sone
$$
where $g^{(1)}_\ell=j_\ell\ln z$ as given in Eq.\Imxi.
As may be expected, the terms in Eq.\Sone\ arising from this
imaginary part just cancel the $j_\ell(\ln z)^2/2$ term of
$X^{(2)}_\ell$ in Eq.\Xxirel. An easy way to check this is to
insert it into the left-hand side of Eq.\PNexpand.
Hence keeping in mind the regularity condition again, we find
$$
 \Re\left(\xi^{(2)}_\ell\right)=f^{(2)}_\ell
   =\cases{O(z^\ell)+O(z^\ell)\ln z
                     -{1\over2}j_\ell(\ln z)^2~&for $\ell=2,3$,\crr
           O(z^{\ell-2})+O(z^\ell)\ln z
                     -{1\over2}j_\ell(\ln z)^2~&for $\ell\geq4$.\cr}
\eqn\Rexitwo
$$

Inserting Eqs.\Rexione\ and \Rexitwo\ into Eq.\Xxirel, we find
$$
\eqalign{
 X^{in}_2
   &=z^3\left[O(1)+\epsilon O(z)
          +\epsilon^2\left\{O(1)+O(1)\ln z\right\}+\cdots\right],
\cr
 X^{in}_3
   &=z^3\left[O(z)+\epsilon O(1)
          +\epsilon^2\left\{O(z)+O(z)\ln z\right\}+\cdots\right],
\cr
 X^{in}_\ell
   &=z^3\left[O(z^{\ell-2})+\epsilon O(z^{\ell-3})
          +\epsilon^2\left\{O(z^{\ell-4})
                   +O(z^{\ell-2})\ln z\right\}
                    +\cdots\right]~(\ell\geq4).
\cr}
\eqn\Xinzero
$$
One sees that $X^{(n)}_\ell$ behaves more regularly than what may be
expected from the boundary condition we discussed in \S2.
As noted previously, in the case of a circular orbit,
we have $z=O(v)$ and $\epsilon=O(v^3)$.
Then from Eq.\Xinzero, we find that $O(\epsilon^3)$ terms for $\ell=2$
and $4$ may contribute to the waveform at $O(v^8)$ beyond Newtonian,
if the leading terms of them behave as $z^{-1}$ relative to the
Newtonian term. By explicitly calculating the series expansion
formulas for $\ell=2$ and $4$, we find it is the case for
$\ell=2$ but not for $\ell=4$; the leading term of order $z^{-1}$
relative to Newtonian cancels out for $\ell=4$.
For completeness, we give the leading term of $X^{(3)}_2$ here:
$$
 X^{(3)}_2={319\over6300}z^2+O(z^4)+O(z^4)\ln z\,.
\eqn\Xthree
$$
Thus in order to
evaluate the gravitational waveform and luminosity
to $O(v^8)$ beyond Newtonian,
it is necessary to know the series expansions of $f^{(n)}_\ell$
for $\ell\leq10-2n$ with $n=0,1,2$ and the leading term of
$f^{(3)}_2$. Note, however, that calculations of these series
expansion formulas are straightforward and easy to obtain.
Note also that, as can be seen from Eq.\dEdt, if we are
interested in the luminosity alone,
we only need to know $f^{(n)}_\ell$ for $\ell\leq6-n$ up to $n=2$
and the leading term of $f^{(3)}_2$.

Recalling again the general formula for the luminosity, Eq.\dEdt,
and the symmetry of $\xi^{(n)}_\ell$, Eq.\xisym,
we see that the $\ln z$ terms in the behaviors of $X^{in}_2$ and
$X^{in}_3$ should give rise to $\ln v$ terms in the luminosity
at $O(v^6)$ and $O(v^8)$ but not at $O(v^7)$ (note that the
$\ln z$ appearing in the above expressions is, mathematically
speaking, actually $\ln|z|$).
This confirms the correctness of the result obtained by Tagoshi
and Nakamura\rlap,\refmark\TagNak at least qualitatively.
Although it may be too early to conclude anything,
the present consideration indicates that the $\ln v$ terms in
the luminosity originate from some local curvature effect near
the source (local in the spatial sense; it may well be non-local
in the temporal sense), but is not due to the effect of curvature
during propagation. This last point will be discussed further below.

Now we turn to the asymptotic behavior at $z=\infty$.
For this purpose, let the asymptotic form of $f^{(n)}_\ell$ be
$$
 f^{(n)}_\ell\rightarrow P^{(n)}_\ell j_\ell+Q^{(n)}_\ell n_\ell
 \quad{\rm as}~z\rightarrow\infty\quad(n=1,2).
\eqn\PQdef
$$
Then noting Eq.\Imxi\ and the equality
$e^{-i\epsilon\ln(z-\epsilon)}=e^{-iz^*}e^{iz}$,
the asymptotic form of $X^{in}_\ell$ is expressed as
$$
\eqalign{
 X^{in}_\ell
   &=e^{-i\epsilon\ln(z-\epsilon)}z
     \biggl[j_\ell+\epsilon\left\{f^{(1)}_\ell-ij_\ell\ln z\right\}
\cr
   &\qquad +\epsilon^2\Bigl\{f^{(2)}_\ell+i\Bigl(-{1\over z}j_\ell
                                       +f^{(1)}_\ell\ln z\Bigr)\Bigr\}
    +\cdots\biggr]
\crr
   &\rightarrow
   {1\over2}e^{-iz^*}\left(zh^{(2)}_\ell e^{iz}\right)
      \biggl[1+\epsilon\left\{P^{(1)}_\ell
                       +i\left(Q^{(1)}_\ell+\ln z\right)\right\}
\cr
   &\qquad
    +\epsilon^2\left\{\left(P^{(2)}_\ell-Q^{(1)}_\ell\ln z\right)
               +i\left(Q^{(2)}_\ell+P^{(1)}_\ell\ln z\right)\right\}
    +\cdots\biggr]
\cr
  &~+{1\over2}e^{iz^*}\left(zh^{(1)}_\ell e^{-iz}\right)
      e^{-2i\epsilon\ln(z-\epsilon)}
      \biggl[1+\epsilon\left\{P^{(1)}_\ell
                       -i\left(Q^{(1)}_\ell-\ln z\right)\right\}
\cr
   &\qquad
    +\epsilon^2\left\{\left(P^{(2)}_\ell+Q^{(1)}_\ell\ln z\right)
                -i\left(Q^{(2)}_\ell-P^{(1)}_\ell\ln z\right)\right\}
    +\cdots\biggr],
\cr}
\eqn\AsymptXin
$$
where $h^{(1)}_\ell$ and $h^{(2)}_\ell$ are the spherical Hankel
functions of the first and second kinds, respectively,
which are given by
$$
 h^{(1)}_\ell=j_\ell+in_\ell\rightarrow (-i)^{\ell+1}{e^{iz}\over z}\,,
\quad
 h^{(2)}_\ell=j_\ell-in_\ell\rightarrow i^{\ell+1}{e^{-iz}\over z}\,.
\eqno\eq
$$
Using the above asymptotic behavior of $h^{(1)}_\ell$ and
$h^{(2)}_\ell$, the incident amplitude $A^{in}_\ell$ can be
readily extracted out:
$$
\eqalign{
  &X^{in}_\ell\rightarrow
       A^{in}_\ell e^{-i(z^*-\epsilon\ln\epsilon)}
       +A^{out}_\ell e^{i(z^*-\epsilon\ln\epsilon)}\,;
\crr
   &A^{in}_\ell={1\over2}i^{\ell+1}e^{-i\epsilon\ln\epsilon}
      \Bigl[1+\epsilon\left\{P^{(1)}_\ell
                       +i\left(Q^{(1)}_\ell+\ln z\right)\right\}
\cr
   &\qquad
    +\epsilon^2\left\{\left(P^{(2)}_\ell-Q^{(1)}_\ell\ln z\right\}
            +i\left(Q^{(2)}_\ell+P^{(1)}_\ell\ln z\right)\right\}
    +\cdots\Bigr],
\cr}
\eqn\Ainlogform
$$
where note that
$$
 \omega r^*=\omega\left(r+2M\ln{r-2M\over2M}\right)
           =z^*-\epsilon\ln\epsilon,
\eqn\zrstar
$$
from our definition of $z^*$, which
gives rise to the phase $-i\epsilon\ln\epsilon$ of $A^{in}_\ell$.
Since we have required $X^{in}_\ell$ to be real, we must have
$A^{out}_\ell=\overline{A^{in}_\ell}$.
This can be explicitly checked by expanding in powers of $\epsilon$
the factor $e^{-2i\epsilon\ln(z-\epsilon)}$ associated with the
terms proportional to $e^{iz^*}$ in Eq.\AsymptXin.

An important point to be noted in the above expression for
$A^{in}_\ell$ is that it contains $\ln z$-dependent terms.
Since $A^{in}_\ell$ should be constant,
$P^{(n)}_\ell$ and $Q^{(n)}_\ell$ should contain appropriate
$\ln z$-dependent terms which exactly cancel the $\ln z$-dependent
terms in the formula \Ainlogform. To be explicit, we must have
$$
\eqalign{
 P^{(1)}_\ell&=p^{(1)}_\ell\,,
\cr
 Q^{(1)}_\ell&=q^{(1)}_\ell-\ln z\,,
\cr
 P^{(2)}_\ell&=p^{(2)}_\ell+q^{(1)}_\ell\ln z-(\ln z)^2\,,
\cr
 Q^{(2)}_\ell&=q^{(2)}_\ell-p^{(1)}_\ell\ln z\,,
\cr}
\eqn\PQlog
$$
where $p^{(n)}_\ell$ and $q^{(n)}_\ell$ ($n=1,2$) are constant.
These relations, as well as Eq.\Imxi\ mentioned previously,
can be used to check the result of calculations.

In terms of $p^{(n)}_\ell$ and $q^{(n)}_\ell$,
$A^{in}_\ell$ is finally expressed as
$$
 A^{in}_\ell
  ={1\over2}i^{\ell+1}e^{-i\epsilon\ln\epsilon}
    \left[1+\epsilon\left(p^{(1)}_\ell+iq^{(1)}_\ell\right)
          +\epsilon^2\left(p^{(2)}_\ell+iq^{(2)}_\ell\right)
          +\cdots\right].
\eqn\Ainform
$$
Note that the above form of $A^{in}_\ell$
implies that the tail radiation, which is due to the curvature
scattering of waves, will contain $\ln v$ terms
as phase shifts in the waveform, but will not give rise to
such terms in the luminosity formula. This supports our previous
argument on the origin of the $\ln v$ terms in the luminosity.

Since $\epsilon=O(v^3)$, the accuracy of $A^{in}_\ell$
to $O(\epsilon^2)$ is sufficient for calculations up to $O(v^8)$
beyond Newtonian.
Specifically, taking account of Eq.\Xinzero\
and the remarks that follow it, we conclude that in order to
calculate waveforms to $O(v^8)$, the accuracy
of $A^{in}_\ell$ we have to achieve for a given $\ell$ is
$O(\epsilon^n)$ where $n=\left[10-\ell\over3\right]$.
On the other hand, if we concentrate on the luminosity,
the required order is $n=\left[12-2\ell\over3\right]$.
Thus, in order to derive the luminosity formula accurate to $O(v^8)$,
we need to evaluate $A^{in}_\ell$ for only $\ell=2,3$ to
$O(\epsilon^2)$ and for $\ell=4$ to $O(\epsilon)$,
but to obtain the waveform to $O(v^8)$ we additionally
need $A^{in}_\ell$ for $\ell=4$ to $O(\epsilon^2)$
and for $\ell=5$ to $O(\epsilon)$. From the observational point of view,
however, the post-Newtonian
terms in the luminosity formula are far more important than
those in the waveform because of their accumulative effect
on the orbital evolution of compact binaries\rlap.\refmark\Cutler\

\chapter{The ingoing-wave Regge-Wheeler function to $O(\epsilon^2)$}

In this section we derive closed analytic formulas of the
ingoing-wave Regge-Wheeler function for arbitrary $\ell$ to
$O(\epsilon)$ and those for $\ell=2,3$ and $4$ to $O(\epsilon^2)$.

\section{Calculation to $O(\epsilon)$}

{}From Eq.\Genform\ with $\xi^{(0)}_\ell=j_\ell$, we have
$$
 \xi_\ell^{(1)}
    =j_\ell I_{\ell,n}^{(1)}-n_\ell I_{\ell,j}^{(1)}
    +{1\over z}j_\ell +ij_\ell\ln z\,.
\eqn\xione
$$
Although a bit tedious, the integrals $I_{\ell,n}^{(1)}$ and
$I_{\ell,j}^{(1)}$ can be straightforwardly evaluated by using
integral formulas for the spherical
Bessel functions. Some useful formulas are given in Appendix.
In addition, the fact that $j_\ell$ and $n_\ell$ appear in an
anti-symmetric manner in Eq.\xione\ helps us to save our
 computational effort considerably.
We then find
$$
\eqalign{
 \xi^{(1)}_\ell
  =&{(\ell-1)(\ell+3)\over2(\ell+1)(2\ell+1)}j_{\ell+1}
    -\left({\ell^2-4\over2\ell(2\ell+1)}
        +{2\ell-1\over\ell(\ell-1)}\right)j_{\ell-1}
\crr
 &~+z^2(n_\ell j_0-j_\ell n_0)j_0
   +\sum_{m=1}^{\ell-2}\left({1\over m}+{1\over m+1}\right)
           z^2(n_\ell j_m-j_\ell n_m)j_m
\crr
 &~+n_\ell\left(\Ci2z-\gamma-\ln 2z\right)-j_\ell\Si2z
       +ij_\ell\ln z+\alpha_\ell^{(1)}j_\ell+\beta_\ell^{(1)}n_\ell\,,
\cr
}\eqn\xionesol
$$
where $\Si x=\int_0^xdt\sin t/t$ and
$\Ci x=-\int_x^\infty dt\cos t/t$ are the sine and cosine
integral functions, respectively,
and $\alpha_\ell^{(1)}$ and $\beta_\ell^{(1)}$
are integration constants to be determined from the boundary
condition. Note that terms like $z^2(n_\ell j_m-j_\ell n_m)$ in
the second line are polynomials in $1/z$ which are known as
the Lommel polynomials:
$$
\eqalign{
 z^2(n_\ell j_m-
   &j_\ell n_m)=R_{\ell-m-1,m+{3\over2}}(z),
\crr
  =&\sum_{r=0}^{\left[\scriptstyle{\ell-m-1\over2}\right]}
     (-1)^r{(\ell-m-1-r)!\,\Gamma(\ell+{1\over2}-r)\over
            r!\,(\ell-m-1-2r)!\,\Gamma(m+{3\over2}+r)}
     \left({2\over z}\right)^{\ell-m-1-2r}.
\cr
}
\eqn\Lommel
$$
As we discussed in \S2, the boundary condition
is that $z\xi^{(1)}_\ell$ is regular at $z=0$. This fixes
$\beta_\ell^{(1)}=0$. As for $\alpha_\ell^{(1)}$, since
it simply represents the arbitrariness of the normalization of
$X^{in}_\ell$, we set $\alpha_\ell^{(1)}=0$ for convenience
(in Poisson's notation\rlap,\refmark{\Poisson} our choice
corresponds to his $\alpha_\ell=-\pi/2$).
Note that this choice is in accordance with the choice we made
in the previous general discussion, as seen by comparing
the asymptotic behavior of $\xi^{(1)}_\ell$ at $z=0$ and Eq.\Rexione.

Asymptotic behavior of $\xi^{(1)}_\ell$ at $z=\infty$
is easily seen. Noting the fact $j_{\ell+1}\sim-j_{\ell-1}\sim
(-1)^{\ell+n}n_{2n-\ell}$, etc., we find
$$
 \xi^{(1)}_\ell\rightarrow
    \left(p^{(1)}_\ell+i\ln z\right)j_\ell
            +\left(q^{(1)}_\ell-\ln z\right)n_\ell\,;
\eqn\xioneinf
$$
where
$$
\eqalign{
  p^{(1)}_\ell=&-{\pi\over2}\,,
\cr
  q^{(1)}_\ell=&{1\over2}\left[\psi(\ell)+\psi(\ell+1)
                   +{(\ell-1)(\ell+3)\over\ell(\ell+1)}\right]-\ln2\,.
\cr}
\eqn\pqoneell
$$
Here $\psi(\ell)$ is the digamma function:
$$
 \psi(\ell)=\sum_{m=1}^{\ell-1}{1\over m}-\gamma\,,
$$
and $\gamma=0.57721\cdots$ is the Euler constant.
Note that Eq.\xioneinf\ is perfectly
consistent with general discussion given in \S3.
$A^{in}_\ell$ to this order is readily found to be
$$
 A^{in}_\ell={1\over2}i^{\ell+1}e^{-i\epsilon\ln\epsilon}
     \left[1+\epsilon\left(-{\pi\over2}+iq^{(1)}_\ell\right)+
      \cdots \right],
\eqn\Ainone
$$
with $q^{(1)}_\ell$ given by Eq.\pqoneell.

Before closing this subsection, we mention an important issue
associated with the interpretation of Eq.\Ainone.
Comparing our result for $\ell=2$ with Eq.(4.24) of Poisson's
paper\rlap,\refmark\Poisson we see that his $\ln\epsilon$
term ($\epsilon=2M\omega$) at $O(\epsilon)$ is automatically
absorbed into the phase of $A^{in}_\ell$ in our result.
An inspection of Poisson's analysis reveals that his $\ln\epsilon$
term indeed originates from the phase of $X^{in}_\ell$
when matching $r^*$ and $r$; $r^*=r+2M\ln[(r-2M)/2M]$,
in the asymptotic behavior of $X^{in}_\ell$;
$e^{\pm i\omega r^*}=e^{\pm i\left(z^*-\epsilon\ln\epsilon\right)}$.
Thus, as discussed in the previous section, this $\ln\epsilon$
term will never contribute to the amplitude of $A^{in}_\ell$ but
only to its phase at any higher order of the expansion.
Then since a different choice of a constant scale in the logarithmic
term of $r^*$ as $2M\rightarrow2M'$ induces only a change of phase by
$i\epsilon\ln(M/M')$, the $\ln\epsilon$ term itself has no
physical significance except for its $\ln\omega$-dependence.

The above argument implies that one should not take
the value of $q^{(1)}_\ell$ itself as physically significant,
since it may also be absorbed into the re-definition of $r^*$ in the
phase of $X^{in}_\ell$.
The physical significance lies in the relative differences
of $q^{(1)}_\ell$ between different $\ell$.
In passing, we note that Wiseman found a discrepancy in the
tail term of the waveform between his result
of a post-Newtonian analysis\refmark\Wiseman
and Poisson's result, which corresponds to
a discrepancy in the value of $q^{(1)}_2$. Wiseman argues that
this discrepancy may be due to different boundary conditions
in the two approaches; with and without the black-hole horizon.
Because of the reason we mentioned above, in addition to
general consideration on the presence of the horizon we gave in \S2,
we suspect Wiseman's argument is incorrect.

\section{Calculation to $O(\epsilon^2)$}

Using recurrence relations for the spherical Bessel functions,
one can reduce all the terms in $\xi^{(1)}_\ell$ of
Eq.\xionesol\ to a linear sum of the spherical Bessel functions,
except for the terms proportional to $\Si2z$, $\Ci2z$ and $\ln z$.
Hence if the integrals involving these latter terms can be evaluated
when inserted into Eq.\Genform, we may obtain an analytic formula
for $\xi^{(2)}_\ell$. Performing integration by parts and utilizing
the anti-symmetric appearance of $j_\ell$ and $n_\ell$,
we then find this condition translates to the condition that
the following integrals are analytically calculable:
$$
\eqalign{
 B_n(z)
    :=&\int_0^z dx\left(2xn_0^2(x)\CC(x)-2xj_0(x)n_0(x)\SS(x)\right),
\cr
 B_j(z)
    :=&\int_0^z dx\left(2xn_0(x)j_0(x)\CC(x)-2xj_0^2(x)\SS(x)\right),
\cr}
\eqn\Bdef
$$
where for notational convenience we have defined
the functions $\CC$ and $\SS$ as
$$
\eqalign{
\CC(z)
  :=&\int_0^{z}dx\left (-2xj_0(x)^2\right)
        =\int_0^{2z}dx{\cos x-1\over x}=\Ci2z-\gamma-\ln 2z\,,
\cr
\SS(z)
  :=&\int_0^{z}dx\left(-2xn_0(x)j_0(x)\right)
      =\int_0^{2z}dx{\sin x\over x}=\Si2z\,.
\cr}
\eqn\CCSS
$$
We note that $\CC(z)=O(z^2)$ and $\SS(z)=O(z)$ so that $B_\zeta(z)$
are perfectly regular at $z=0$.
Now using the fact $n_0^2(z)=-j_0^2(z)+1/z^2$ and
the above definitions of $\CC(z)$ and $\SS(z)$,
we can rewrite $B_\zeta(z)$ ($\zeta=n,j$) as
$$
\eqalign{
 B_n(z)
  =&\int_0^z dx\,\left(\CC'(x)\CC(x)+\SS'(x)\SS(x)\right)
    +2\int_0^z {dx\over x}\CC(x)
\cr
  =&{1\over2}\left(\CC(z)^2+\SS(z)^2\right)+2\CC(z)\ln 2z
        +2\int_0^{2z}{dx\over x}(1-\cos x)\ln x\,,
\crr
 B_j(z)
  =&\int_0^z dx\,\left(\CC'(x)\SS(x)-\SS'(x)\CC(x)\right)
\cr
  =&(\gamma-\ln2z)\SS(z)+2\int_0^{2z}{dx\over x}\sin x\ln x
\cr
  &\qquad\quad
   +\int_0^{2z}{dx\over x}\left(\Si x\cos x-\Ci x\sin x\right)\,.
\cr}
\eqn\Bnj
$$
Hence the asymptotic behavior of $B_\zeta$ at $z=\infty$ can be
obtained if the integrals in the above expressions can be evaluated
for $z\rightarrow\infty$.
In Appendix we show this can be done:
$$
\eqalign{
 &2\int_0^{2z}{dx\over x}\sin x\ln x
  \rightarrow-\pi\gamma\,,
\crr
 &2\int_0^{2z}{dx\over x}(1-\cos x)\ln x
  \rightarrow (\ln2z)^2+{\pi^2\over12}-\gamma^2,
\crr
 &\int_0^{2z}{dx\over x}\left(\Si x\cos x-\Ci x\sin x\right)
  \rightarrow0.
\cr}
\eqn\Bnjint
$$
Using these results, we find
$$
\eqalign{
 &B_n\rightarrow{5\over24}\pi^2-{1\over2}(\gamma+\ln2z)^2,
\crr
 &B_j\rightarrow-{\pi\over2}(\gamma+\ln2z)\,,
\cr}
\eqn\Binfty
$$
as $z\rightarrow\infty$.

Thus, it is possible to derive an analytic formula of
$\xi^{(2)}_\ell$ for arbitrary $\ell$,
though the manipulation will be quite involved.
However, as discussed in the previous section, as long as we are
concerned with the accuracy up to $O(v^8)$ beyond Newtonian,
it is unnecessary to derive a general formula of $\xi^{(2)}_\ell$.
Instead, at $O(\epsilon^2)$, we only need the series expansion
formulas of $\xi^{(2)}_\ell$ at $z=0$ for $\ell\leq6$ and the
asymptotic formulas of $\xi^{(2)}_\ell$ at $z=\infty$ for $\ell\leq4$.
The former can be easily obtained by expanding in powers of $z$
the integrands in Eq.\Genform.
Hence it is sufficient to derive closed analytic formulas of
$\xi^{(2)}_\ell$ up to $\ell=4$ for calculation of the
luminosity and waveforms with accuracy to $O(v^8)$.
In what follows, we present the results of our calculation for
$\ell=2,3$ and $4$ in order.
\endpage
\medskip
\leftline{(a) \undertext{\strut$\ell=2$}}
\smallskip

First consider the case of $\ell=2$.
{}From Eq.\Genform, we have
$$
\eqalign{
 f^{(2)}_2
   =&\Re\left(\xi^{(2)}_2\right)
\crr
   =&{1\over z}f^{(1)}_2
    +j_2\int dz(2zn_2-n_3)f^{(1)}_2-n_2\int dz(2zj_2-j_3)f^{(1)}_2
\crr
   &+j_2\int dz\,n_2j_2-n_2\int dz\,j_2^2-{1\over2}j_2(\ln z)^2,
\crr
 g^{(2)}_2
  =&\Im\left(\xi^{(2)}_2\right)
\crr
   =&{1\over z}j_2\ln z
    +j_2\int dz(2zn_2-n_3)j_2\ln z-n_2\int dz(2zj_2-j_3)j_2\ln z
\crr
   &-j_2\int dz\,z\left(n_2f^{(1)}_2{}'-f^{(1)}_2n_2{}'\right)
    +n_2\int dz\,z\left(j_2f^{(1)}_2{}'-f^{(1)}_2j_2{}'\right).
\cr}
\eqn\xitwoform
$$
Calculation of the above integrals is straightforward but
requires care and patience.
As discussed in \S2, the imaginary part $g^{(2)}_2$
should be equal to the one given in Eq.\Imxi. With the boundary
condition that it be regular at $z=0$, we have explicitly
evaluated the above integral expression for $g^{(2)}_2$ and
confirmed our expectation.

As for the real part $f^{(2)}_2$, we obtain
$$
\eqalign{
  f^{(2)}_2
  =&-{389\over70z^2}j_0-{113\over420z}j_1+{1\over7z}j_3
     +B_n(z) j_2-B_j(z) n_2
\crr
  &+\CC(z)\left(
   {5\over6z}n_2-{5\over3}n_1-{3\over z}n_0-{107\over210}j_2\right)
\crr
  &-\SS(z)\left(
   {5\over6z}j_2-{5\over3}j_1-{3\over z}j_0+{107\over210}n_2\right)
\crr
  &-{107\over210}j_2\ln z-{1\over2}j_2(\ln z)^2+\alpha^{(2)}_2j_2\,,
\cr}
\eqn\ftwosol
$$
where $\CC$, $\SS$, $B_n$ and $B_j$ are the functions
defined in Eqs.\Bdef\ and \CCSS, and $\alpha^{(2)}_2$
is an arbitrary constant which we shall set to zero for convenience.
For confirmation, we have examined
the above expression for $f^{(2)}_2$
with the help of the algebraic manipulation program, {\it Mathematica}.
We have confirmed that it satisfies the basic equation
\PNexpand\ and the required boundary condition at $z=0$, Eq.\Rexitwo.

Asymptotic behavior of $f^{(2)}_2$ at $z=\infty$ is now
easy to evaluate. Again, we find the behavior in agreement
with general discussion of \S3:
$$
\eqalign{
 f^{(2)}_2
  &\rightarrow P^{(2)}_2j_2+Q^{(2)}_2n_2\,;
\crr
  &P^{(2)}_2=p^{(2)}_2+q^{(1)}_2\ln z-(\ln z)^2\,,
\cr
  &Q^{(2)}_2=q^{(2)}_2-p^{(1)}_2\ln z\,,
\cr}
\eqno\eq
$$
where $p^{(1)}_2$ and $q^{(1)}_2$ can be read off from the
general formula \pqoneell\ in the previous subsection,
$$
 p^{(1)}_2=-{\pi\over2}\,,\quad q^{(1)}_2={5\over3}-\gamma-\ln2\,,
\eqno\eq
$$
and the constants $p^{(2)}_2$ and $q^{(2)}_2$ are found to be
$$
\eqalign{
 p^{(2)}_2
    &=-{1\over2}\left(q^{(1)}_2\right)^2
      +{1\over2}\left({5\over3}\right)^2+{5\over24}\pi^2
            +{107\over210}(\gamma+\ln2),
\cr
 q^{(2)}_2
   &=-{\pi\over2}\left(q^{(1)}_2+{107\over210}\right).
\cr}
\eqn\pqtwo
$$
Hence $A^{in}_2$ to $O(\epsilon^2)$ is found to be
$$
\eqalign{
 A^{in}_2
  =&-{1\over2}ie^{-i\epsilon\ln\epsilon}
   \biggl[1+\epsilon\left\{-{\pi\over2}+iq^{(1)}_2\right\}
\cr
   &\qquad+\epsilon^2
       \biggl\{-{1\over2}\left(q^{(1)}_2\right)^2
             +{25\over18}+{5\over24}\pi^2+{107\over210}(\gamma+\ln2)
\cr
   &\qquad\qquad
      -i{\pi\over2}\left(q^{(1)}_2+{107\over210}\right)\biggr\}
            +\cdots\biggr].
\crr
  =&-{1\over2}ie^{-i\epsilon(\ln2\epsilon+\gamma)}
     \exp\left(i\epsilon{5\over3}-i\epsilon^2{107\over420}\pi\right)
\crr
  \times&\biggl(1-\epsilon{\pi\over2}
  +\epsilon^2
       \biggl\{{25\over18}+{5\over24}\pi^2+{107\over210}(\gamma+\ln2)
       \biggr\}+\cdots\biggr),
\cr}
\eqn\Aintwo
$$
where, in the last expression, we have absorbed the terms arising
from $q^{(1)}_2$ and $q^{(2)}_2$ into the phase.

\medskip
\leftline{(b) \undertext{\strut $\ell=3$}}
\smallskip

Now we turn to the case of $\ell=3$. We focus on the
real part of $\xi^{(2)}_3$.
Then Eq.\Genform\ yields
$$
\eqalign{
 f^{(2)}_3
   =&\Re\left(\xi^{(2)}_3\right)
\cr
   =&{1\over z}f^{(1)}_3
    +j_3\int dz(2zn_3-{12\over7}n_4-{5\over7}n_2)f^{(1)}_3
\cr
   &\qquad -n_3\int dz(2zj_3-{12\over7}j_4-{5\over7}j_2)f^{(1)}_3
\cr
   &\qquad+j_3\int dz\,n_3j_3-n_3\int dz\,j_3^2-{1\over2}j_3(\ln z)^2.
\cr}
\eqn\xithrform
$$
As before, after a lengthy calculation,
we obtain
$$
\eqalign{
 f^{(2)}_3
  =&{1\over4z}j_4+{323\over49z}j_2-{5065\over294z^2}j_1
    -\left({1031\over588z}+{445\over14z^3}\right)j_0
    +{65\over6z^2}n_0-{65\over6z}n_1
\crr
 &+\CC(z)\left({3\over2z}n_3-{13\over6}n_2-{9\over2z}n_1
       -{15\over z^2}n_0-{13\over42}j_3\right)
\crr
 &-\SS(z)\left({3\over2z}j_3-{13\over6}j_2-{9\over2z}j_1
       -{15\over z^2}j_0+{13\over42}n_3\right)
\crr
 &+B_n(z)j_3-B_j(z)n_3
      -{13\over42}j_3\ln z-{1\over2}j_3(\ln z)^2+\alpha^{(2)}_3j_3\,,
\cr}
\eqn\fthree
$$
where $\alpha^{(2)}_3$ is an arbitrary constant which
will be put to zero hereafer.
We have also examined the above result with {\it Mathematica\/} and
confirmed it.

Asymptotic behavior of $f^{(2)}_3$ at $z=\infty$ is
$$
\eqalign{
 f^{(2)}_3
  &\rightarrow P^{(2)}_3j_3+Q^{(2)}_3n_3\,;
\crr
  &P^{(2)}_3=p^{(2)}_3+q^{(1)}_3\ln z-(\ln z)^2\,,
\cr
  &Q^{(2)}_3=q^{(2)}_3-p^{(1)}_3\ln z\,,
\cr}
\eqno\eq
$$
where $p^{(2)}_3$ and $q^{(2)}_3$ are found to be
$$
\eqalign{
 p^{(2)}_3
    &=-{1\over2}\left(q^{(1)}_3\right)^2
     +{1\over2}\left({13\over6}\right)^2+{5\over24}\pi^2
            +{13\over42}(\gamma+\ln2),
\crr
 q^{(2)}_3
   &=-{\pi\over2}\left(q^{(1)}_3+{13\over42}\right),
\cr}
\eqn\pqthree
$$
with $p^{(1)}_3$ and $q^{(1)}_3$, again from Eq.\pqoneell,
being given by
$$
 p^{(1)}_3=-{\pi\over2}\,,\quad q^{(1)}_3={13\over6}-\gamma-\ln2\,.
\eqno\eq
$$
By absorbing the $q^{(1)}_3$ and $q^{(2)}_3$ terms into the phase,
$A^{in}_3$ to $O(\epsilon^2)$ is found to be
$$
\eqalign{
 A^{in}_3
  =&{1\over2}e^{-i\epsilon(\ln2\epsilon+\gamma)}
    \exp\left(i\epsilon{13\over6}-i\epsilon^2{13\over84}\pi\right)
\crr
  \times&\biggl(1-\epsilon{\pi\over2}
  +\epsilon^2
       \biggl\{{169\over72}+{5\over24}\pi^2+{13\over42}(\gamma+\ln2)
        \biggr\}+\cdots\biggr).
\cr}
\eqn\Ainthree
$$

\medskip
\leftline{(c) \undertext{\strut$\ell=4$}}
\smallskip

The calculation of $\xi^{(2)}_4$ is parallel to the previous
two cases.
{}From Eq.\Genform, we have
$$
\eqalign{
 f^{(2)}_4
   =&\Re\left(\xi^{(2)}_4\right)
\cr
   =&{1\over z}f^{(1)}_4
    +j_4\int dz(2zn_4-{7\over3}n_5-{4\over3}n_3)f^{(1)}_4
\cr
   &\qquad -n_4\int dz(2zj_4-{7\over3}j_5-{4\over3}j_3)f^{(1)}_4
\cr
   &\qquad+j_4\int dz\,n_4j_4-n_4\int dz\,j_4^2-{1\over2}j_4(\ln z)^2.
\cr}
\eqn\xifourform
$$
Then we obtain
$$
\eqalign{
f^{(2)}_4=
 &{56\over 165\,z}j_5+
   \left(-{{5036}\over {33\,{z^4}}}+{30334\over 1155\,z^2}\right)j_4
   +\left({35252\over 33\,z^5}-{30334\over 165\,z^3}
   +{14401\over 3465\,z}\right)j_3
\crr
 &-\left({5036\over 11\,z^5} + {{45461}\over {693\,{z^3}}}
         +{{36287}\over {9240\,z}}\right)j_1
  +\left({{140}\over {{z^3}}} - {{5}\over {18\,z}}\right)n_0
  -{{49}\over {6\,z}}n_2
\crr
 &+C(z)\,\left({21\over10\,z}n_4 -{149\over60}n_3+{5\over3\,z}n_2
    -{105\over2\,z^2}n_1-{105\over z^3}n_0+{10\over z}n_0
    -{1571\over6930}j_4 \right)
\crr
 &-S(z)\,\left({21\over10\,z}j_4-{149\over60}j_3+{5\over3\,z}j_2
    -{105\over 2\,z^2}j_1-{105\over z^3}j_0+{10\over z}j_0
    +{1571\over6930}n_4 \right)
\crr
 &+B_n(z)\,j_4 - B_j(z)\,n_4 - {1571\over6930}\,j_4\,\ln z
  -{1\over2}\,j_4\,(\ln z)^2+\alpha^{(2)}_4j_4\,,
\cr}
\eqn\ffour
$$
where, for convenience, we shall choose $\alpha^{(2)}_4=0$
also in this case.

{}From the asymptotic behavior of $f^{(2)}_4$ at $z=\infty$,
we find
$$
\eqalign{
 p^{(2)}_4
    &=-{1\over2}\left(q^{(1)}_4\right)^2
       +{1\over2}\left({149\over60}\right)^2+{5\over24}\pi^2
            +{1571\over6930}(\gamma+\ln2),
\crr
 q^{(2)}_4
   &=-{\pi\over2}\left(q^{(1)}_4+{1571\over6930}\right),
\cr}
\eqn\pqfour
$$
where $p^{(1)}_4$ and $q^{(1)}_4$ are once again
read off from Eq.\pqoneell\ as
$$
 p^{(1)}_4=-{\pi\over2}\,,\quad q^{(1)}_4={149\over60}-\gamma-\ln2\,.
\eqno\eq
$$
Hence $A^{in}_4$ to $O(\epsilon^2)$ is obtained as
$$
\eqalign{
 A^{in}_4
  =&{1\over2}ie^{-i\epsilon(\ln2\epsilon+\gamma)}
    \exp\left(i\epsilon{149\over60}
              -i\epsilon^2{1571\over13860}\pi\right)
\crr
  \times&\biggl(1-\epsilon{\pi\over2}
  +\epsilon^2
       \biggl\{{22201\over7200}+{5\over24}\pi^2
      +{1571\over6930}(\gamma+\ln2)
        \biggr\}+\cdots\biggr).
\cr}
\eqn\Ainfour
$$

\chapter{Conclusion}

We have formulated a new post-Newtonian type expansion to
solve the homogeneous Regge-Wheeler equation.
We have then applied it to solve for the ingoing-wave Regge-Wheeler
function $X^{in}_\ell$, which is purely ingoing at the black-hole
horizon. We have focused on $X^{in}_\ell$ since it plays an
essential role in evaluating the gravitational waveform and
luminosity radiated to infinity by a particle orbiting around a
non-rotating black hole.
The method has an advantage that the ingoing-wave boundary condition
of $X^{in}_\ell$ at the horizon is naturally taken into account.
It also allows a systematic expansion of $X^{in}_\ell$ in powers
of $\epsilon=2M\omega$, which is of $O(v^3)$ where $v$ is the
orbital velocity of a particle.

By investigating general features of $X^{in}_\ell$,
we have argued that as long as we are concerned with the radiation
going out to infinity, the effect of the radiation absorbed into
the black-hole horizon is negligible until we go to $O(\epsilon^6)$,
or $O(v^{18})$ beyond Newtonian. Hence $X^{in}_\ell$
can be also used for a situation in which there is
a non-rotating compact star instead of a black hole.
Thus our method can provide an important guideline for the
study of inspiraling compact binaries, though it is restricted
to the case when one body has a small mass relative to the other.

Then we have solved for $X^{in}_\ell$ and derived its analytic
expression correct to $O(\epsilon)$ for arbitrary $\ell$
and to $O(\epsilon^2)$ for $\ell=2,3$ and $4$. These results can be
used to analytically evaluate the gravitational waveform and the
luminosity radiated by a particle in circular orbit
to $O(v^8)$ beyond Newtonian.
This is in progress and the result will be published in a
forthcoming paper\rlap.\refmark\TagSas\

\ack
I would like to thank T. Nakamura and H. Tagoshi for stimulative
conversations and explanation of their semi-analytic results given in
Ref.\TagNak), by which I was prompted to study the present subject.
This work was supported in part by the Grant-in-Aid
for Scientific Research on Priority Areas of
Ministry of Education, Science and Culture, No.04234104.
\par\penalty-300\vskip\chapterskip
   \spacecheck\chapterminspace \chapterreset \xdef\chapterlabel{A}
   \titlestyle{Appendix}\nobreak\vskip\headskip \penalty 30000
   \wlog{\string\Appendix\ \chapterlabel}

In this Appendix, we summarize mathematical formulas associated
with the spherical Bessel functions which are used in the calculation
of the ingoing-wave function $X^{in}_\ell$.
In doing so, we also sketch the calculation
to $O(\epsilon^2)$.
We denote any of the spherical Bessel functions by $\zeta_m$
and/or by $\zeta_m^*$.

The basic recurrence relation is
$$
 \zeta_{m-1}+\zeta_{m+1}={2m+1\over z}\zeta_m\,,
\eqn\recur
$$
from which it follows that
$$
  j_{-m}=(-1)^m\,n_{m-1}
    \quad\leftrightarrow\quad
  n_{-m}=(-1)^{m-1}\,j_{m-1}\,.
\eqn\nandj
$$
Thus all the spherical Bessel functions appearing in the text
can be expressed in terms of $j_m$ alone if necessary.
There is also a useful series expansion formula:
$$
 j_{m+n}=\left({z\over2}\right)^n
            \sum_{r=0}^{n}{n!\,(m+2r+1/2)\,\Gamma(m+r+1/2)\over
                          r!\,(n-r)!\,\Gamma(m+n+r+3/2)}
            j_{m+2r}\quad (n\geq0).
\eqn\jseries
$$
This enables us to get rid of all the inverse powers of $z$
appearing, e.g., in the integrands of Eq.\xitwoform. Hence
except for the terms involving $\Ci2z$, $\Si2z$ and $\ln z$,
all the terms in the integrands reduce to either of
the form $z\zeta_m\zeta_n^{*\mathstrut}$ or of the form
$\zeta_m\zeta_n^{*\mathstrut}$.
For such forms, the following indefinite integral formulas
are known to hold:
$$
\eqalign{
 &\int dz\,\zeta_{m}\zeta_{n}^{*\mathstrut}
  ={z^2\over(m-n)(m+n+1)}
      \left(\zeta_{m}\zeta_{n+1}^{*\mathstrut}
         +\zeta_{m-1}\zeta_{n}^{*\mathstrut}\right)
      -{z\over m-n}\zeta_{m}\zeta_{n}^{*\mathstrut}\quad (m\neq n),
\crr
 &\int dz\,\zeta_\ell\zeta_\ell^{*\mathstrut}
  ={1\over2\ell+1}\left\{
     \int dz\,\zeta_0\zeta_0^{*\mathstrut}
         -z\biggl(\zeta_0\zeta_0^{*\mathstrut}
         +2\sum_{m=1}^{\ell-1}\zeta_{m}\zeta_{m}^{*\mathstrut}
         +\zeta_\ell\zeta_\ell^{*\mathstrut}
         \biggr)\right\},
\crr
 &\int dz\,z\zeta_{m}\zeta_{n}^{*\mathstrut}
  =\int dz\,z\zeta_{m-1}\zeta_{n-1}^{*\mathstrut}-{z^2\over m+n}
      \left(\zeta_{m-1}\zeta_{n-1}^{*\mathstrut}
        +\zeta_{m}\zeta_{n}^{*\mathstrut}\right),
\crr
 &\int dz\,z\zeta_\ell\zeta_\ell^{*\mathstrut}
  =\int dz\,z\zeta_0\zeta_0^{*\mathstrut}-{z^2\over2}\left\{
     \zeta_0\zeta_0^{*\mathstrut}+\sum_{m=1}^{\ell-1}
        \left({1\over m}+{1\over m+1}\right)
           \zeta_{m}\zeta_{m}^{*\mathstrut}
         +{1\over\ell}\zeta_\ell\zeta_\ell^{*\mathstrut}\right\},
\cr}
\eqn\Intform
$$
and
$$
\eqalign{
 &\int dz\, n_0^2 =-\int dz\,j_0^2-{1\over z}=-\SS(z)-zn_0^2,
\crr
 &\int dz\, n_0j_0=-\CC(z)-zn_0j_0-\ln z\,,
\crr
 &\int dz\, zn_0^2=-\int dz\, zj_0^2+\ln z={1\over2}\CC(z)+\ln z\,,
\crr
 &\int dz\, zn_0j_0=-{1\over2}\SS(z),
\cr}
\eqn\Intzero
$$
where the functions $\SS(z)$ and $\CC(z)$ are defined in Eq.\CCSS,
which we recapitulate:
$$
\SS(z)=\Si2z\,,\qquad \CC(z)=\Ci2z-\gamma-\ln 2z\,.
\eqn\CCSSap
$$

As for the terms involving $\CC(z)$, $\SS(z)$ and $\ln z$,
what we need to evaluate are the asymptotic forms of the
integrals in the expressions of $B_n(z)$ and $B_j(z)$ given
by Eq.\Bnj:
$$
\eqalign{
  I_{s}(z)
    :=&\int_0^{2z}{dx\over x}\sin x\ln x\,,
\cr
  I_{c}(z)
    :=&\int_0^{2z}{dx\over x}(1-\cos x)\ln x\,,
\cr
  I_{cs}(z)
    :=&\int_0^{2z}{dx\over x}\left(\Si x\cos x-\Ci x\sin x\right)\,.
\cr}
\eqn\Icscs
$$
First, let us evaluate $I_s$. It can be done as
$$
\eqalign{
 I_s(z)
   &={\partial\over\partial\alpha}
       \left(\int_0^{2z}dxx^{\alpha-1}\sin x\right)_{\alpha=0}
\crr
   &\Buildrel{z\rightarrow\infty}\under\longrightarrow
     {\partial\over\partial\alpha}
       \left(\Gamma(1+\alpha)
        {\sin{\pi\over2}\alpha\over\alpha}\right)_{\alpha=0}
\crr
   &=-{\pi\over2}\gamma.
\cr}
\eqn\Isinf
$$
In a similar manner, $I_c$ is evaluated as
$$
\eqalign{
 I_c(z)
   &={\partial\over\partial\alpha}
       \left(\int_0^{2z}dxx^{\alpha-1}(1-\cos x)\right)_{\alpha=0}
\crr
   &={\partial\over\partial\alpha}
       \left({(2z)^\alpha-1\over\alpha}
         -\left[\int_0^{2z}dxx^{\alpha-1}\cos x-{1\over\alpha}\right]
            \right)_{\alpha=0}
\crr
   &\Buildrel{z\rightarrow\infty}\under\longrightarrow
    {1\over2}(\ln2z)^2-
     {\partial\over\partial\alpha}
       \left({\Gamma(1+\alpha)\cos{\pi\over2}\alpha-1
                  \over\alpha}\right)_{\alpha=0}
\crr
   &={1\over2}(\ln2z)^2+{\pi^2\over24}-{1\over2}\gamma^2.
\cr}
\eqn\Icinf
$$
Finally, we consider $I_{cs}$. We utilize the following
integral representation of the integrand:
$$
 \si x\cos x-\ci x\sin x=-\int_0^\infty dt{\sin t\over t+x}\,,
\eqno\eq
$$
where $\si x=\Si x-\pi/2$ and $\ci x=\Ci x$.
Then we manipulate the integral as
$$
\eqalign{
 I_{cs}
  &=\int_0^{2z}dx\left\{
   \int_0^\infty dt{\sin t\over t}
       \left({1\over t+x}-{1\over x}\right)
             +{\pi\over2}{\cos x\over x}\right\}
\crr
 &=\int_0^\infty {dt\over t}\sin t\int_0^{2z}{dx\over t+x}
   +{\pi\over2}\int_0^{2z}dx{\cos x-1\over x}
\crr
 &=\int_0^\infty{dt\over t}\sin t\ln(2z+t)
    -\int_0^\infty{dt\over t}\sin t\ln t +{\pi\over2}\CC(z)
\crr
 &={\pi\over2}\left(\ln2z+\gamma+\CC(z)\right)
      +\int_0^\infty{dt\over t}\sin t\ln\left(1+{t\over2z}\right)
\crr
 &={\pi\over2}\Ci2z
      +\int_0^\infty{dt\over t}\sin 2zt\,\ln(1+t)
 \Buildrel{z\rightarrow\infty}\under\longrightarrow0.
\cr}
\eqno\eq
$$
Hence $I_{cs}$ vanishes as $z\rightarrow\infty$.

\refout
\end